\documentclass{emulateapj}

\journalinfo{To appear in the Astrophysical Journal}
\submitted{Received March 30, 2009; Accepted August 20, 2009.}

\shorttitle{An Interferometric Study of the Fomalhaut Inner Debris Disk. I.}
\shortauthors{Absil et al.}

\begin{document}


\title{An interferometric study of the Fomalhaut inner debris disk \\ I. Near-infrared detection
of hot dust with VLTI/VINCI\footnotemark[$\ast$]} \footnotetext[$\ast$]{Based on observations made with ESO Telescopes at the Paranal Observatory (public VINCI commissioning data).}


\author{Olivier Absil\altaffilmark{1}}
\affil{IAGL, Universit\'e de Li\`ege, 17 All\'ee du Six Ao\^ut, B-4000 Sart Tilman, Belgium}
\affil{LAOG--UMR 5571, CNRS and Universit\'e Joseph Fourier, BP 53, F-38041 Grenoble, France}

\author{Bertrand Mennesson}
\affil{Jet Propulsion Laboratory, California Institute of Technology, 4800 Oak Grove Drive,
Pasadena, CA 91109, USA}

\author{Jean-Baptiste Le Bouquin}
\affil{European Southern Observatory, Casilla 19001, Santiago 19, Chile}

\author{Emmanuel Di Folco and Pierre Kervella}
\affil{LESIA--UMR 8109, CNRS and Observatoire de Paris-Meudon, 5 place J.~Janssen, F-92195 Meudon, France}

\and

\author{Jean-Charles Augereau}
\affil{LAOG--UMR 5571, CNRS and Universit\'e Joseph Fourier, BP 53, F-38041 Grenoble, France}

\altaffiltext{1}{FNRS Postdoctoral Researcher}



\begin{abstract}
The innermost parts of dusty debris disks around main sequence stars are currently poorly known due to the high contrast and small angular separation with their parent stars. Using near-infrared interferometry, we aim to detect the signature of hot dust around the nearby A4\,V star Fomalhaut, which has already been suggested to harbor a warm dust population in addition to a cold dust ring located at about 140\,AU. Archival data obtained with the VINCI instrument at the VLTI are used to study the fringe visibility of the Fomalhaut system at projected baseline lengths ranging from 4\,m to 140\,m in the $K$ band. A significant visibility deficit is observed at short baselines with respect to the expected visibility of the sole stellar photosphere. This is interpreted as the signature of resolved circumstellar emission, producing a relative flux of $0.88\% \pm 0.12\%$ with respect to the stellar photosphere. While our interferometric data cannot directly constrain the morphology of the excess emission source, complementary data from the literature allow us to discard an off-axis point-like object as the source of circumstellar emission. We argue that the thermal emission from hot dusty grains located within 6\,AU from Fomalhaut is the most plausible explanation for the detected excess. Our study also provides a revised limb-darkened diameter for Fomalhaut ($\theta_{\rm LD} = 2.223 \pm 0.022$\,mas), taking into account the effect of the resolved circumstellar emission.
\end{abstract}




\section{Introduction}

\setcounter{footnote}{1}

The young \citep[$\sim$\,200\,Myr,][]{DiFolco04} and nearby (7.7\,pc) A4 main sequence star Fomalhaut ($\alpha$~PsA, HD~216956) has been the focus of much attention during the last decade in the context of planetary system studies. Although the discovery of a cold debris disk around this bright star dates back to early observations with the InfraRed Astronomical Satellite \citep{Aumann85}, the first resolved observations of its disk have been obtained only in the late 90's in the sub-millimeter regime \citep{Holland98}, and a few years later at optical wavelengths \citep{Kalas05}. Fomalhaut has nowadays one of the best studied debris disk, with most of its (cold) dust arranged in a narrow ring at 140\,AU from its host star. This ring shows interesting features, such as a sharp inner edge that has been interpreted as the result of the gravitational influence of a massive planet located just inside the dust ring \citep{Quillen06}. The existence of this predicted planetary companion was recently confirmed by direct observations of Fomalhaut in the optical regime \citep{Kalas08}, further boosting the general interest of this system. The Fomalhaut debris disk has additionally been suggested to contain a population of warm dust thanks to partially resolved Spitzer/MIPS observations and Spitzer/IRS spectroscopy \citep{Stapelfeldt04}.

Although the Fomalhaut debris disk has been studied in many details during the last few years, its inner dust content remains rather elusive, with only weak constraints on a warm dust population within 20\,AU provided by Spitzer observations. The main challenges for characterizing this warm dust population are the small angular separation and the high contrast between the star and the inner disk. Infrared interferometry is an appropriate tool to tackle these challenges: by providing an angular resolution as good as a few milli-arcseconds (mas), it can potentially resolve dust populations down to a fraction of an AU from Fomalhaut. This paper is the first of a series that aims at using infrared interferometry to characterize the warm dust content in the inner few AUs around Fomalhaut, and thereby provide a better view on the global architecture of its planetary system.

The present paper focuses on the search for hot circumstellar emission in the near-infrared regime with the VINCI instrument of the VLT Interferometer. The principle for warm dust detection with interferometry is based on the fact that the stellar photosphere and its surrounding dust disk have different spatial scales. For an A-type star at 7.7\,pc, the angular diameter of the photosphere is about 2\,mas, while the circumstellar disk extends beyond the sublimation radius of dust grains, typically located around 0.15\,AU (i.e., 20\,mas) for black body grains sublimating at $T_{\rm sub}\simeq 1500$\,K. The circumstellar disk is thus fully resolved at baselines as short as 10\,m in the near-infrared and contributes as an incoherent background emission at longer baselines, while the stellar photosphere is only resolved at about 150\,m. The presence of the circumstellar disk then shows up as a decrease of visibility at baselines longer than about 10\,m with respect to the expected visibility of the stellar photosphere. This detection method, which works best at short baselines where the stellar photosphere is completely unresolved, has already been successfully used to detect circumstellar emission around the main sequence stars Vega \citep{Absil06}, $\tau$~Cet \citep{DiFolco07}, $\zeta$~Aql \citep{Absil08} and $\beta$~Leo \citep{Akeson09}. In this paper, we revisit archival VLTI/VINCI observations of Fomalhaut and apply the same method to obtain first evidence for hot dust within 6\,AU from Fomalhaut.


\section{Observations and Data Reduction}

The bright star Fomalhaut has been observed on several occasions between 2001 and 2004 with VINCI, the VLT Interferometer Commissioning Instrument, which coherently combines the infrared light
coming from two telescopes in the infrared $H$ and $K$ bands \citep[for a detailed description, see][]{Kervella03b}. Some of these observations have already been reported in two papers by \citet{DiFolco04} and \citet{LeBouquin06}. In this section, we describe these two data sets as well as an unpublished data set that we have extracted from the ESO archives. Most of the observations described here have been obtained with the 40-cm test siderostats of the VLTI. Their field-of-view (FOV) is limited by the use of single-mode fibers inside the VINCI instrument, and can be described by a 2D Gaussian function with a full width at half maximum (FWHM) of $1\farcs6$ in the $K$ band under standard atmospheric conditions at Cerro Paranal. This translates into a linear FOV radius at half maximum of about 6\,AU at the distance of Fomalhaut.

    \subsection{Long-baseline data from \citet{DiFolco04}}

\begin{figure*}
\plotone{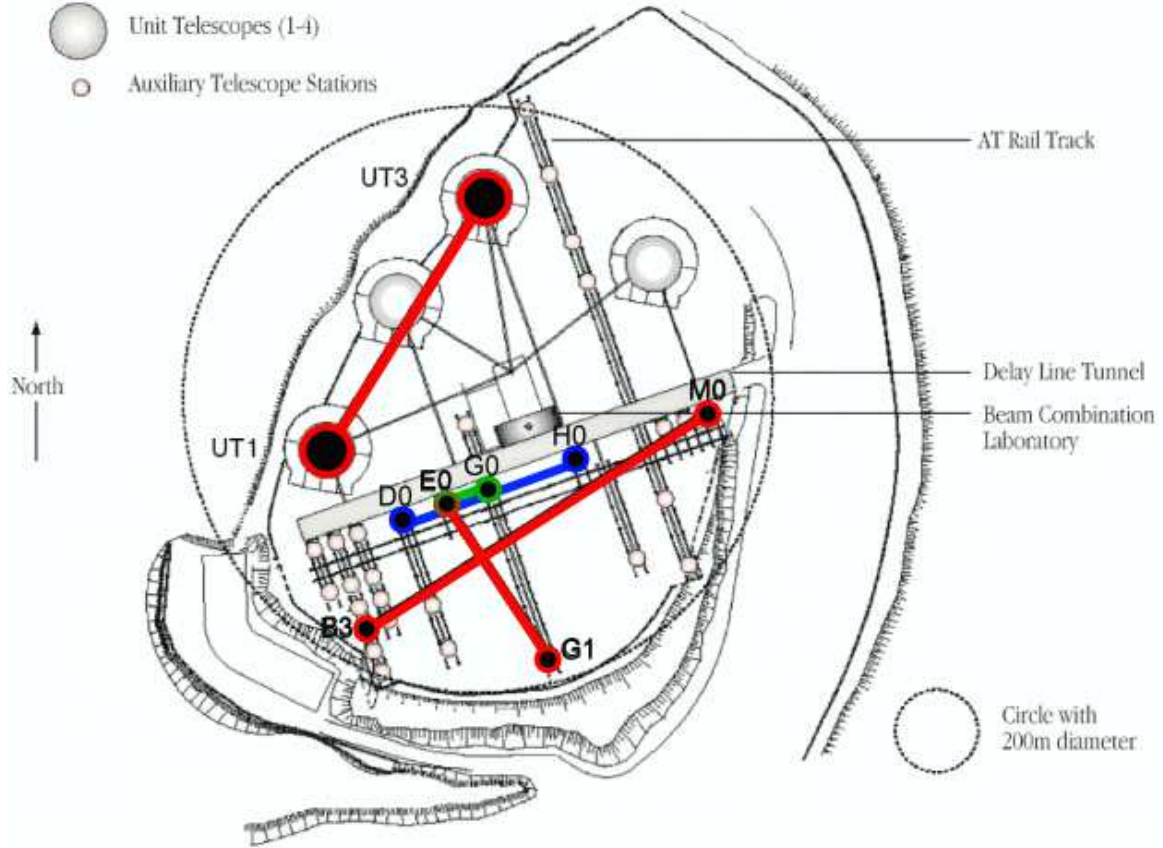}
\caption{Schematic view of the VLTI baselines used in this study. Data at long baselines (UT1--UT3, B3--M0, E0--G1; in red) and intermediate baselines (D0--H0, in blue) are taken from previous studies respectively by \citet{DiFolco04} and \citet{LeBouquin06}, while the data obtained at the shortest baselines (E0--G0, in green) are unpublished data from the ESO archives.}
\label{fig:vlti}
\end{figure*}

Observations of Fomalhaut have been obtained in the $K$ band by \citet{DiFolco04} between November 2001 and October 2002 with three baselines of various lengths and orientations: E0--G1 (66\,m) and
B3--M0 (140\,m) with the 40\,cm test siderostats, and UT1--UT3 (102\,m) with the 8\,m Unit Telescopes (see Fig.~\ref{fig:vlti}). A few observations have also been obtained in the $H$ band on the E0--G1 baseline, but because $H$-band data are not available on other baselines and therefore cannot be used to assess the presence of circumstellar emission in this particular waveband, these observations will not be further discussed here.

In the present study, we will use the calibrated squared visibilities published by \citet{DiFolco04}. Data reduction was performed by using the wavelet analysis of the fringe power spectrum described by \citet{Kervella04} and implemented in the VNDRS data reduction software\footnote{Publicly available on the ESO VLTI web site (see {\tt http://www.eso.org/projects/vlti/instru/vinci/drs/}).}. The resulting coherence factors (or raw squared visibilities) have been converted into calibrated visibilities by estimating the instrumental transfer function with interleaved observations of calibrator stars with known diameters. The uncertainties on the resulting squared visibilities have been separated into statistical error (related to the dispersion of the raw visibilities) and systematic error (induced by the uncertainty on the calibrators' angular diameters).

This data set has allowed \citet{DiFolco04} to derive an accurate estimation of the
uniform-disk angular diameter ($\theta_{\rm UD}$) of Fomalhaut: $2.197 \pm 0.011 \pm 0.020$\,mas. The first error bar refers to the statistical dispersion of the measurements whereas the second takes into account the systematic effects. This UD diameter can be converted into an equivalent limb-darkened diameter ($\theta_{\rm LD}$) following \citet{HanburyBrown74}:
\begin{equation}
\frac{\theta_{\rm LD}(\lambda)}{\theta_{\rm UD}(\lambda)} =
\sqrt{\frac{1-u_{\lambda}/3}{1-7u_{\lambda}/15}} \; , \label{eq:UDLD}
\end{equation}
with $u_{\lambda}$ the linear limb-darkening coefficient at a given wavelength $\lambda$. Using the tabulated limb-darkening coefficients of \citet{Claret00}, the resulting LD diameter reads $2.228 \pm 0.011 \pm 0.020$\,mas. Due to the lack of short baselines, \citet{DiFolco04} could not thoroughly study the circumstellar environment of Fomalhaut, and were only able to derive an upper limit of about $3\%$ on the flux ratio between the circumstellar and photospheric emissions.

    \subsection{Medium-baseline data from \citet{LeBouquin06}} \label{sub:dataJB}

Observations of Fomalhaut have been obtained with VINCI in the $K$ band using two 40\,cm
siderostats on the 64\,m D0--H0 baseline (see Fig.~\ref{fig:vlti}). The observations were performed on five consecutive nights from October 8th to October 12th, 2004. A large part of this data set was reduced by \citet{LeBouquin06} using a custom data reduction procedure, and was subsequently used to validate the new integrated optics beam-combiner IONIC-2TK, which was installed in replacement of the original MONA fibered beam-combiner in August 2004. The study performed by \citet{LeBouquin06} shows a very nice consistency between the MONA and IONIC-2TK versions of the VINCI instrument, including very similar spectral transmissions. The study also demonstrates the validity of a Fourier-based estimator to reduce the IONIC-2TK data.

These observations were not intended to study the circumstellar environment of Fomalhaut, so that
the observing strategy was not optimal in that respect. In particular, only one calibrator star
(88~Aqr) was used throughout the five nights. This is expected to produce significant systematic
errors on the calibrated visibilities, all the more that, with its estimated angular diameter of $3.240 \pm 0.057$\,mas \citep{Richichi09}, this calibrator star is partly resolved on the D0--H0 baseline. However, because some of the observations have been obtained for projected baseline lengths as short as 20\,m, this data set is still very useful to assess the presence of circumstellar emission. At such baselines, the stellar photosphere is almost unresolved ($V^2 \simeq 0.98$), while the fully resolved circumstellar emission can induce a significant deficit of squared visibility. It must also be noted that, based on this large data set and taking advantage of the high internal stability of VINCI, \citet{LeBouquin06} have derived a UD angular diameter of $3.26 \pm 0.06$\,mas for 88~Aqr, which we will use in this study, as well as an independent estimation of the UD angular diameter of Fomalhaut ($2.23 \pm 0.07$\,mas), in good agreement with the estimation of \citet{DiFolco04}.

\begin{figure}
\plotone{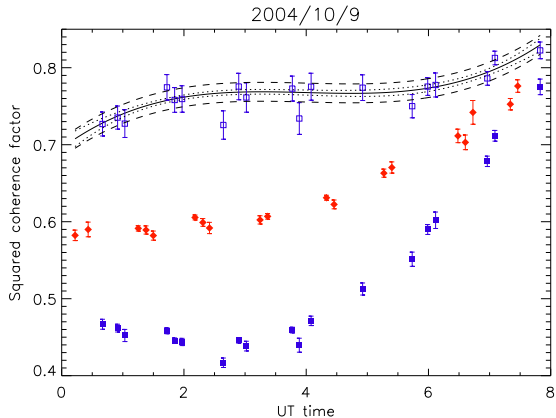}
\caption{Squared coherence factors ($\mu^2$) and interferometric transfer function estimations ($T^2$) for one representative observing night on the D0--H0 baseline with VINCI-IONIC-2TK. Filled symbols correspond to $\mu^2$ data points for the scientific target (Fomalhaut, red diamonds) and the calibrator target (88~Aqr, blue squares). The interferometric transfer function (TF) is represented by the empty blue squares, which are derived from the calibrator measurements taking into account its known diameter. The solid line is an interpolation of the TF, obtained by a third degree polynomial fit to its estimated values. The 1-$\sigma$ statistical and systematic error bars on the estimation of the TF are respectively represented by dotted and dashed lines.}
\label{fig:T2IONIC}
\end{figure}

For the present study, we have retrieved the full original data set (5 nights) and reduced it with VNDRS~3.1, the latest version of the VINCI data reduction software. A total of 86~individual
observations are available on Fomalhaut for the five nights. The output of the software consists in wavelet-based Fourier domain estimations of the squared coherence factors $\mu^2$ for the target star (Fomalhaut) and its calibrator (88~Aqr). These ``uncalibrated visibilities'' are illustrated for a representative night in Fig.~\ref{fig:T2IONIC}. Coherence factors must then be converted into squared visibilities ($V^2$) for the scientific object by the relation:
\begin{equation}
V^2 = \frac{\mu^2}{T^2}
\end{equation}
where $T^2$ is the interferometric transfer function (TF), i.e., the response of the system to a
point source. The estimation of the TF is based on the interleaved observations of the calibrator
star, whose $V^2$ can be computed from the a priori knowledge of its angular diameter ($\theta_{\rm UD} = 3.26 \pm 0.06$\,mas). This $V^2$ must be computed for the wide spectral bandwidth on which the VINCI observations are performed, taking into account the actual spectrum of the star (88~Aqr, a K1\,III giant) and the spectral transmission of the VINCI instrument. The spectral transmission of VINCI is taken from \citet{Kervella03}, and we have used for 88~Aqr the tabulated $K$-band spectrum for a K1\,III star in the \citet{Pickles98} stellar spectral flux library. The interferometric TF is then estimated as $T^2 = \mu^2/V^2$, and is illustrated for one representative night in Fig.~\ref{fig:T2IONIC}. The statistical and systematic error bars on this estimation are computed following Appendix C of \citet{Kervella04}.

Once the interferometric TF has been estimated at the time of the calibrator observations, the
value of the TF and its error bars must be estimated at the time of the actual scientific
observations. We have used two different techniques to perform the interpolation of the TF between
the calibrator data points: a linear interpolation between the two calibrator observations
bracketing a given scientific observations \citep[as in][]{Kervella04}, and a global polynomial fit on the whole night \citep[as in][]{LeBouquin06}. We have checked that the two methods give
identical results within error bars, and decided to use the second one for the rest of the study. A third order polynomial was found appropriate to follow the (small) variations of the
interferometric TF during the whole night, as the target and calibrator moved across the sky. With
this method, the statistical error bar on the interferometric TF is computed locally at any given
time by using the covariance matrix on the parameters of the polynomial fit, which is available on output of the standard IDL routine {\tt poly\_fit.pro}. The systematic error bar, on the other
hand, is computed globally with a weighted sum of the systematic error bars on all $T^2$
estimations, taking into account the correlation between calibrators. In the present case, since
only one calibrator was used throughout the night, the global systematic error is equal to the average of the systematic error bars on all $T^2$ estimations.

    \subsection{Short-baseline data from the ESO archives}

Even though the projected baseline lengths covered by the \citet{LeBouquin06} data set are sufficiently short to study the circumstellar environment of Fomalhaut, we have searched the ESO archives for calibrated VINCI observations on even shorter baselines. We have identified four interesting nights in 2003 (September 27th, September 30th, November 2nd and November 3rd), where observations of Fomalhaut interleaved with a calibrator star (88~Aqr) have been obtained on the 16\,m E0--G0 baseline, using the 40\,cm test siderostats and the MONA beam combiner. We have used VNDRS~3.1 to reduce this data set. One night (November 2nd) was discarded due to poor atmospheric conditions, and the remaining three nights produced 44 individual observations of Fomalhaut of sufficient quality.

\begin{figure}
\plotone{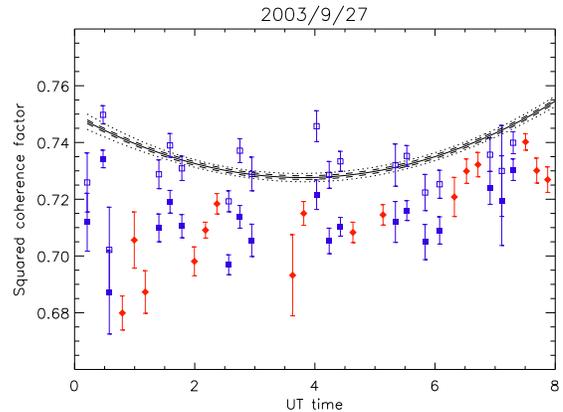}
\caption{Same as Fig.~\ref{fig:T2IONIC} for one representative observing night on the E0--G0 baseline with VINCI-MONA. A second degree polynomial was used to fit the estimated values of the interferometric TF.}
\label{fig:T2MONA}
\end{figure}

Following the method described in Section~\ref{sub:dataJB}, we have used the interleaved observations of the calibrator star 88~Aqr to estimate the interferometric TF. In the present case, a second-degree polynomial was found convenient to model the TF, as illustrated for a representative night in Fig.~\ref{fig:T2MONA}. The scatter in the interferometric TF estimations in Fig.~\ref{fig:T2MONA} may seem larger than in Fig.~\ref{fig:T2IONIC}, but this is actually mostly due to the different scales for the two plots and to the smaller systematic error bars in the present data set. The reduced size of the systematic error bars is related to the fact that the calibrator star is mostly unresolved on the very short E0--G0 baseline, so that the uncertainty on its angular diameter translates into a much smaller systematic error on the TF estimation.

\begin{figure}
\plotone{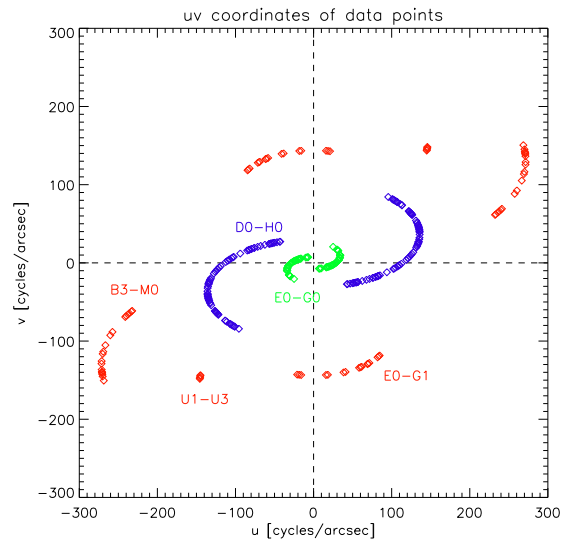}
\caption{Sampling of the Fourier ($u$,$v$) plane for our complete data set, using the same color code as in Fig.~\ref{fig:vlti}.}
\label{fig:uvcoord}
\end{figure}

Our final VINCI data set on Fomalhaut, spanning a large range of baseline lengths and azimuths,
comprises a total of 171 calibrated squared visibilities. The spatial frequencies
sampled by our observations are well spread in the Fourier ($u,v$) plane, as illustrated in
Fig.~\ref{fig:uvcoord}.

\section{Evaluating the amount of circumstellar emission}

\begin{figure*}[t]
\plotone{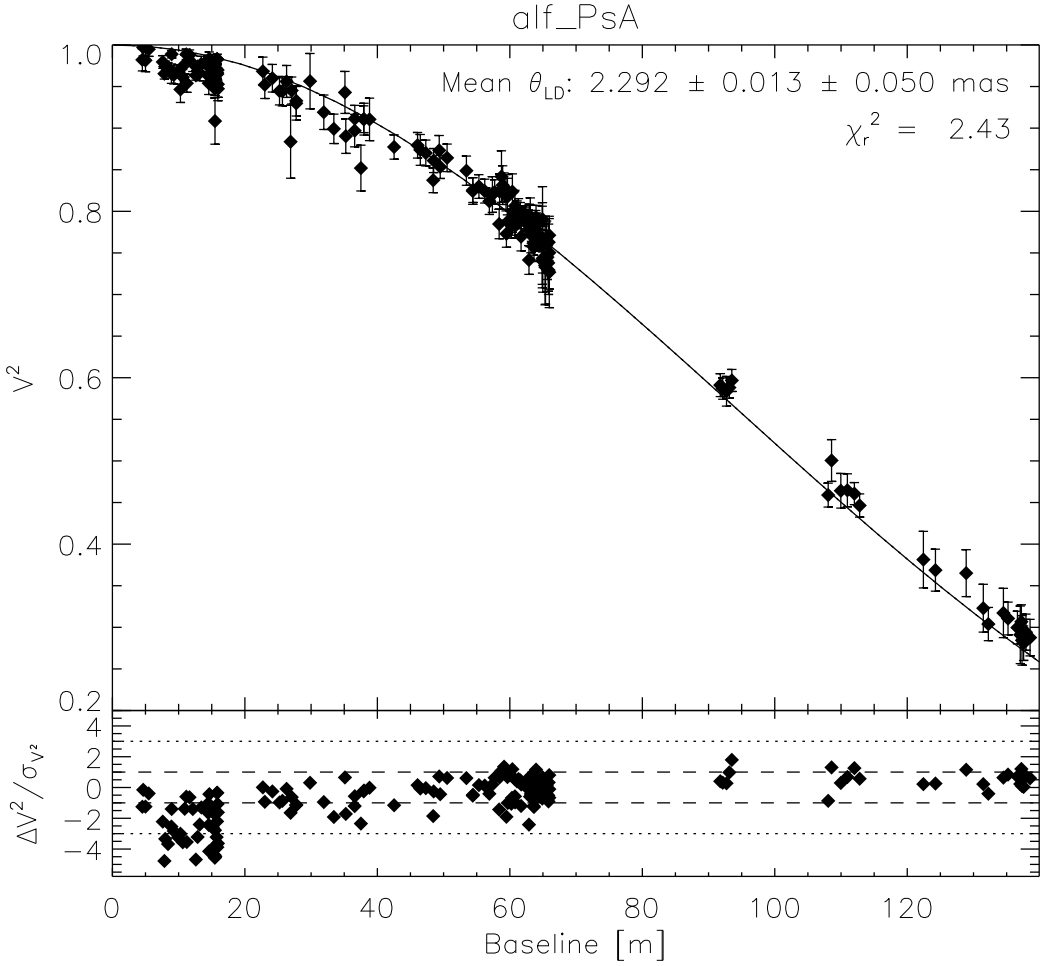}
\caption{Result of the fit of an oblate limb-darkened stellar photosphere to our full data set. For the sake of clarity, the solid line corresponds to a circular photosphere with a diameter equal to the geometric mean of the diameter of the best-fit oblate stellar model, while the residuals are computed with the full 2D photospheric model. The dashed and dotted lines in the bottom panel represent respectively a $1\sigma$ and a $3\sigma$ deviation with respect to the best-fit model.}
\label{fig:fitstar}
\end{figure*}

Let us first assess whether a realistic stellar photospheric model for Fomalhaut could reproduce the whole data set. The description of our oblate limb-darkened photospheric model is given in Appendix~\ref{app:photosphere}, showing that only one parameter needs to be fitted ($\theta_{\rm LD}$, the geometric mean of the limb-darkened diameter). Fitting our model to the whole data set gives $\theta_{\rm LD} = 2.292 \pm 0.013 \pm 0.050$\,mas (Fig.~\ref{fig:fitstar}), where the first error bar accounts for the statistical dispersion of the data while the second error bar is related to the systematic error in the evaluation of the interferometric TF. This result is within $2\sigma$ of the previous estimation by \citet{DiFolco04}, but the residuals of the fit in Fig.~\ref{fig:fitstar} and the large reduced chi square ($\chi_r^2 = 2.43$) clearly demonstrate that an oblate limb-darkened photosphere cannot reproduce the data satisfactorily. It must be noted that our data sample various baseline azimuths, which makes it difficult to represent the model together with the data on a single plot. The solid line of Fig.~\ref{fig:fitstar} is actually computed for a perfectly circular photosphere of diameter $\theta_{\rm LD}$ and is only an approximation of the actual model used to compute the fit residuals, which take into account the various baseline azimuths and the stellar oblateness.

Accepting the photospheric position angle recently measured by \citet{Lebouquin09}, possible improvements of the fit could be obtained by changing either the limb-darkening coefficient ($u_K$) or the photospheric oblateness ($\rho$). This is investigated in Table~\ref{tab:quality}, which gives the goodness-of-fit for various values of these two parameters. The fit is slightly better for extreme limb-darkening coefficients ($u_K \lesssim 1$), and for a slightly more oblate photosphere than expected from our model. But the differences in $\chi_r^2$ are so small that we can safely state that our interferometric data do not constrain the shape of the photosphere in a significant way. We will therefore use the most plausible values (oblateness $\rho=1.021$ and position angle ${\rm PA}=156 \degr$, see Appendix~\ref{app:photosphere}) in the rest of this study.

\begin{deluxetable}{ccccc}[t]
\tablecaption{Goodness-of-fit ($\chi_r^2$) for the photospheric model as a function of oblateness and limb-darkening \label{tab:quality}}
\tablewidth{0pt}
\tablehead{\colhead{} & \colhead{$u_K=0.0$} & \colhead{$u_K=0.19$\tablenotemark{a}} & \colhead{$u_K=0.5$} & \colhead{$u_K=1.0$}}
\startdata
$\rho=1.00$                  & 2.46 & 2.46 & 2.45 & 2.43 \\
$\rho=1.021$\tablenotemark{b} & 2.44 & 2.43 & 2.42 & 2.41 \\
$\rho=1.05$                  & 2.42 & 2.42 & 2.41 & 2.40 \\
$\rho=1.10$                  & 2.46 & 2.46 & 2.45 & 2.44
\enddata
\tablenotetext{a}{Expected value of the limb-darkening coefficient $u_K$ in the $K$ band \citep{Claret00}.}
\tablenotetext{b}{Expected value of the photospheric oblateness $\rho$ based on our oblate photosphere model (see Appendix~\ref{app:photosphere}).}
\end{deluxetable}

    \subsection{Fitting a star-disk model to the complete data set} \label{sub:diskfit}

The failure of a realistic stellar photospheric model to reproduce the interferometric data urges us to increase the complexity of our model. The apparent decrement of visibility observed at short baselines and the slope in the residuals of the fit suggest that another source of emission, fully resolved by the interferometer, must be present within the interferometric field-of-view. As a first step, we assume that this resolved emission is associated with the circumstellar debris disk, and to represent its contribution we use a simple model of a diffuse source uniformly distributed across the whole field-of-view. As discussed by \citet{Absil06}, such a model is a good approximation provided that the circumstellar disk is fully resolved at all baselines. It must be noted that in the present case, due to the very short baselines, this condition might not be completely fulfilled. The uniform emission model should nevertheless provide a good estimation of the flux ratio between the integrated circumstellar emission and the stellar photosphere. More realistic circumstellar emission models will be discussed in Section~\ref{sec:diskmodel}.

\begin{figure*}[t]
\plotone{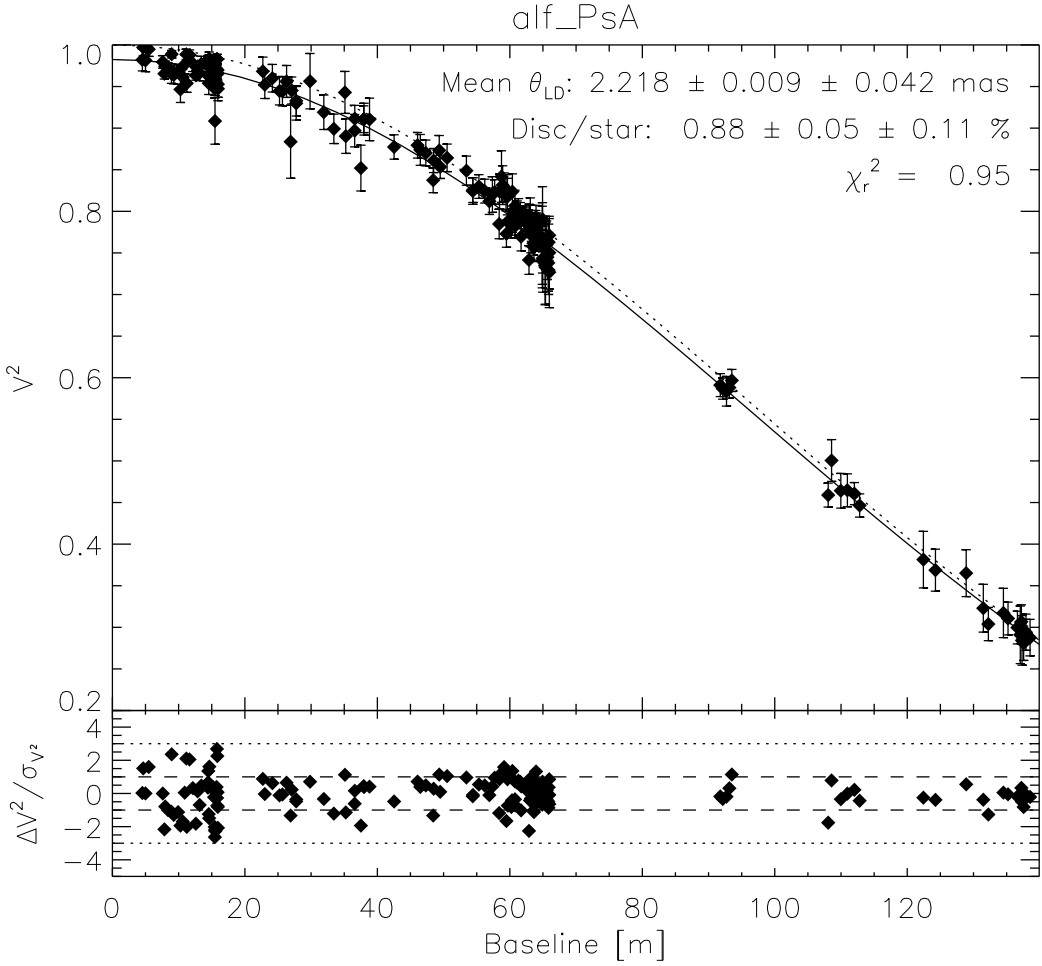}
\caption{Result of the fit of a star-disk model to our full data set. The solid line represents the best fit star-disk model, while the dotted line represents the best-fit result with a single star for comparison, using the same conventions as in Fig.~\ref{fig:fitstar}.}
\label{fig:fitdisk1}
\end{figure*}

In Fig.~\ref{fig:fitdisk1}, we have fitted the whole VINCI data set with a limb-darkened oblate stellar photosphere surrounded by a uniform circumstellar emission. The quality of the fit is very satisfactory, with $\chi_r^2=0.95$ and no obvious trend in the residuals, which are now nicely spread around 0. The best-fit mean limb-darkened diameter is $\theta_{\rm LD} = 2.218 \pm 0.009 \pm 0.042$\,mas, while the best-fit flux ratio between the circumstellar disk and the star is $0.88 \pm 0.05 \pm 0.11$\%. Assuming that our model is representative of the actual brightness distribution, circumstellar emission is thus detected at a $7\sigma$ level. Another representation of our detection is given in Fig.~\ref{fig:fitdisk2} and discussed in Appendix~\ref{app:shortbase}.

\begin{figure*}[t]
\plottwo{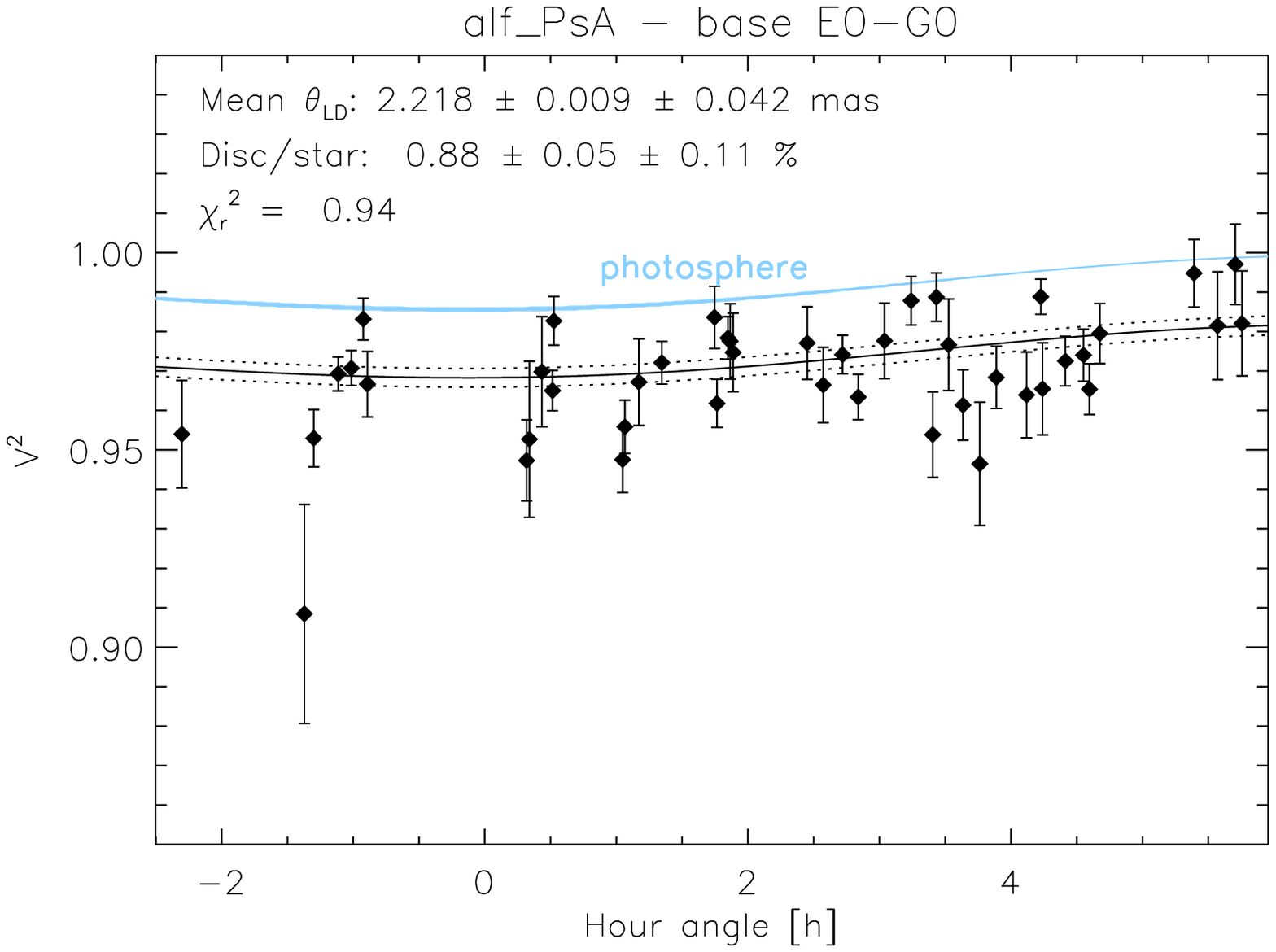}{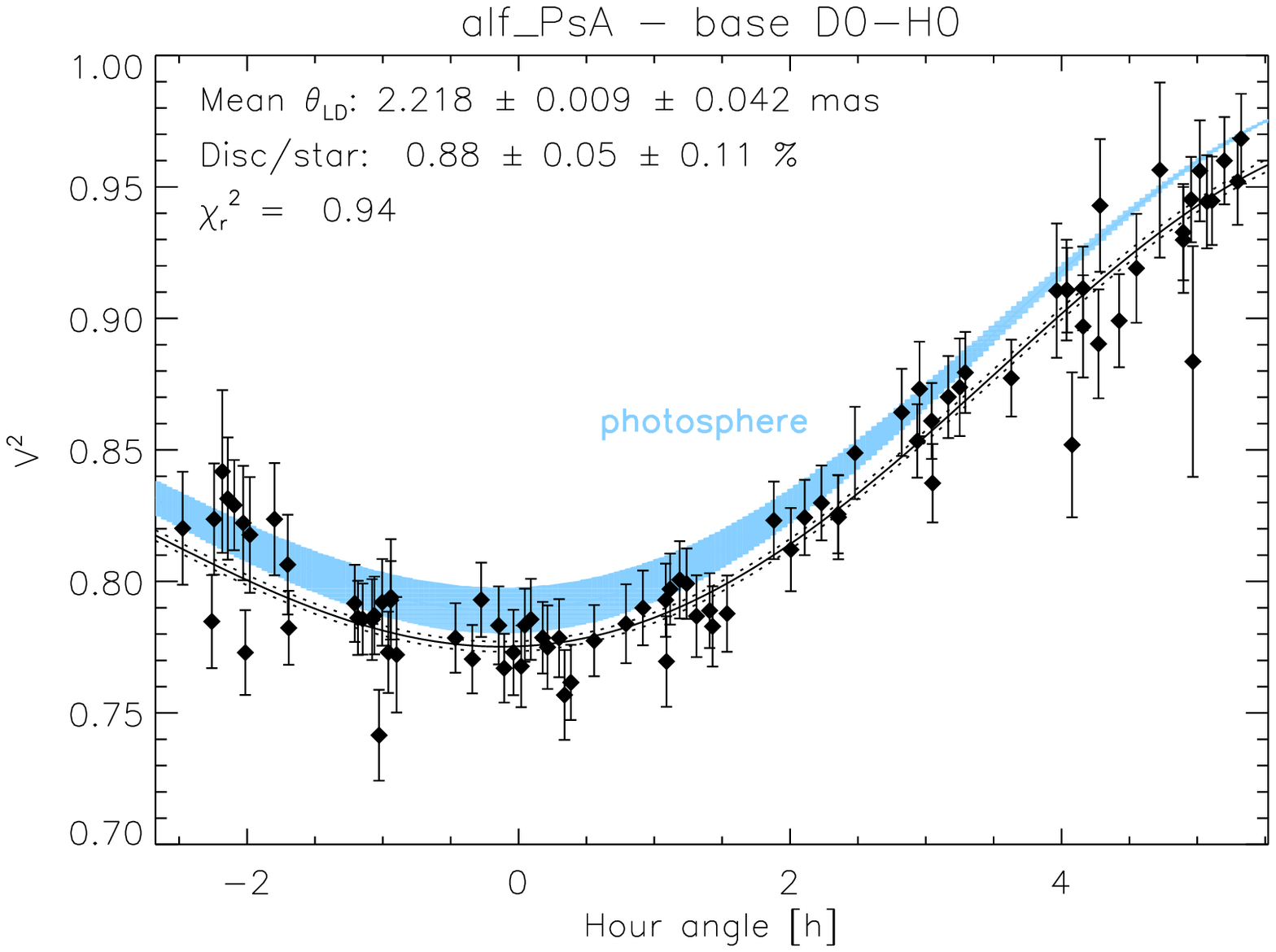}
\caption{Result of the fit of a star-disk model to our full data set, showing the data for two particular baselines as a function of the hour angle of the observation (\textit{left}: the small E0--G0 baseline, \textit{right}: the mid-range D0--H0 baseline). The blue curve represents the visibility of the star alone for our best-fit star-disk model, and its thickness corresponds to the 3$\sigma$ uncertainty related to the final error bar on its diameter. The black solid line represents the best-fit star-disk model, with its 3$\sigma$ confidence interval shown as dotted lines.}
\label{fig:fitdisk2}
\end{figure*}

Systematic errors dominate the noise budget in our detection. This is not surprising as our short- and intermediate-baseline data are calibrated with a single calibrator star (88~Aqr). The uncertainty on this calibrator's diameter is the main contributor to the systematic error budget. Using a few different calibrators would have been preferable to reduce systematic errors, but the archival data were not originally meant to be used for this particular purpose. We are nonetheless confident in the robustness of our result for three main reasons. First, 88~Aqr is a K giant with a slow rotational velocity \citep[3.6\,km\,s$^{-1}$,][]{Hekker07} that has already been used as an infrared photometric standard \citep{Bouchet91} and as an interferometric calibrator by various authors without any hint of unexpectedly high or low visibilities. This includes a recent study dedicated to interferometric calibrators by \citet{Richichi09}, where a UD diameter of $3.240\pm0.057$\,mas is derived. Second, we have independently estimated its angular diameter using surface-brightness relationships \citep{Kervella04b}, giving a UD diameter of $3.31\pm0.08$\,mas in good agreement with the value used in this study ($3.26\pm0.06$\,mas). Finally, we note that an imperfect model for 88~Aqr cannot be at the origin of the observed visibility deficit for the following reasons. Artificially decreasing its angular diameter would increase the calibrated visibilities of Fomalhaut. This would however not reconcile the Fomalhaut calibrated data set with a simple photospheric model, as it would have a much larger influence on the data collected at mid-range baselines (D0--H0) than on the short baseline data (E0--G0). A change in the limb-darkening coefficient would not help either. The presence of a circumstellar environment (disk or companion) around the calibrator could only reduce the measured calibrator visibilities with respect to their expected values, which would result in an increase of the calibrated visibilities for the scientific target and thereby reduce the measured disk/star flux ratio with respect to its actual value. Therefore, the probability that our detection is based on a systematic effect related to the calibrator is considered to be extremely low.

Our revised estimation of the stellar limb-darkened diameter of Fomalhaut ($\theta_{\rm LD} = 2.218 \pm 0.009 \pm 0.042$\,mas) is within the error bar of the previous estimation by \citet{DiFolco04}, as it is mainly based on long-baseline data which are hardly affected by the presence of the circumstellar emission. The large systematic error is due to the fact that most data (at short- and mid-baseline lengths) have been calibrated with the same reference star. A better accuracy on the stellar diameter can actually be obtained by fixing the disk/star contrast at 0.88\% and by fitting only the long-baseline data, which have been obtained with five different calibrators. The final result is then $\theta_{\rm LD} = 2.223 \pm 0.006 \pm 0.021$\,mas ($\chi_r^2=0.31$), which is in very good agreement with the estimation of \citet{DiFolco04}. The actual angular diameter ranges between $2.246\pm 0.022$\,mas and $2.200\pm 0.022$\,mas (apparent major and minor axes), taking into account the apparent oblateness of 1.021.

    \subsection{Fitting a binary star model to the complete data set}

Besides a circumstellar disk, another potential source of visibility deficit at short baselines would be a faint point-like object within the interferometric FOV around the target star. To reproduce the observed visibility drop, the off-axis object should have a flux ratio of about 0.88\% with respect to Fomalhaut as seen through the interferometer in the $K$ band, which is equivalent to a magnitude $K=6.1$ for a close companion. Because of the Gaussian shape of the off-axis coupling efficiency into the single-mode fibers, the companion must be brighter than $K=6.1$ to reproduce the VINCI near-infrared excess if its angular separation is a significant fraction of (or larger than) the radius at half maximum ($0\farcs8$) of the instrumental FOV. Following the discussion in \citet{Absil06}, we estimate that the presence of such a bright background object within the instrumental FOV is very unlikely (probability $\sim10^{-6}$). Therefore we only consider the case of a bound low-mass companion.

\begin{figure*}[t]
\plotone{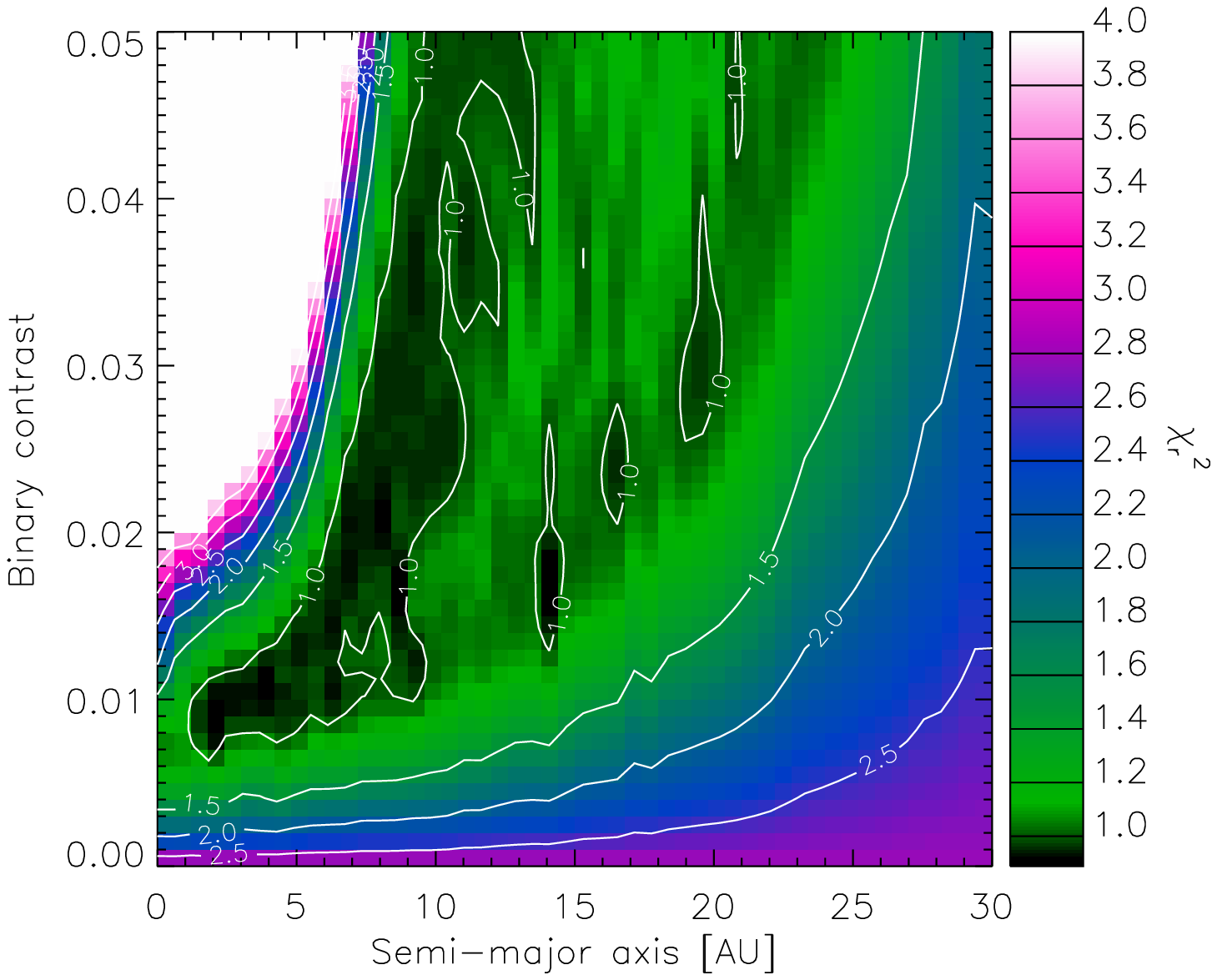}
\caption{Reduced chi square map for the fit of a binary star model, plotted as a function of the orbital semi-major axis of the companion and of the binary flux ratio. The orbital dynamics are taken into account in this model, assuming a circular orbit, and the initial orbital phase is optimized to minimize the $\chi^2_r$. We have limited the scale range to $\chi_r^2 \le 4$ for the sake of clarity. Contours are shown for $\chi^2_r$ between 1 and 3.5, with a step of 0.5.}
\label{fig:chi2binary}
\end{figure*}

To fit our interferometric data set, we construct a model of a binary star, assuming that the companion orbits within the plane defined by the outer dust ring imaged at visible and sub-millimetric wavelengths. We therefore set the inclination and position angle of the projected orbit to $65\fdg9$ and $156\degr$, respectively. In practice, coplanarity with the outer dust ring is probably not mandatory to ensure its stability, but the purpose here is only to check whether binary star models could actually fit the data rather than to explore all the possible orbital solutions. For the sake of simplicity, we will further assume that the orbit is circular, so that we are left with three parameters: the semi-major axis, the orbital phase at a given time $t_0$ and the binary flux ratio (or {\em contrast}). Using a large number of potential values for these three parameters, we compute the position of the companion at our observing dates and the associated visibility of the binary system. We deduce the chi square between the observed visibilities and the computed ones, and for each couple semi-major axis / contrast, we search for the minimum $\chi^2$ as a function of the orbital phase at $t_0$. Thereby, we produce a $\chi^2$ map, which is represented in Fig.~\ref{fig:chi2binary} as a function of semi-major axis and binary contrast.

The $\chi^2$ map shows that a whole range of semi-major axes and contrasts fit the interferometric data in a satisfactory way. The effect of the Gaussian beam profile is evident in this figure, as the binary contrast required to fit the data increases with the binary separation. We have restricted the plot range to contrasts smaller than 5\%, as brighter companions would most probably have been detected by simple near-infrared spectro-photometric measurements. Because the instrumental transmission becomes extremely small at linear radii larger than 30\,AU (i.e., angular separations larger than $4\arcsec$), binaries with larger semi-major axes cannot reproduce our interferometric data set. Conversely, for any semi-major axis smaller than about 25\,AU, one can find suitable combinations of orbital parameters and flux ratio to fit the data satisfactorily (reduced $\chi^2 \sim 1$). For the closest companions, located within the radius at half maximum of the FOV transmission ($0\farcs8$, i.e., about 6\,AU), the best-fit binary flux ratio is about 0.88\% as expected.

In conclusion, even though our interferometric measurements span a large range of time, baseline lengths and azimuths, they are not sufficient to discriminate between a circumstellar disk and a point-like companion as the source of visibility deficit. This is mostly due to the (lack of) observing strategy, which was not optimized for such a goal.


\section{Nature of the near-infrared excess source}

The previous section has shown that our interferometric data set is not sufficient to discriminate between a point-like object and an extended source as the origin of the near-infrared excess emission. Therefore, we need to use complementary data to further constrain the nature of the near-infrared emission around Fomalhaut. In this section, we focus on the possible presence of a low-mass companion, as well as on potential non-thermal sources of extended emission. The thermal emission from a circumstellar dust disk will be briefly discussed in Sections~\ref{sec:diskmodel} and \ref{sec:discussion}.

    \subsection{Further constraints on a low-mass companion}

Since a low-mass companion cannot be ruled out by the interferometric data, we have searched the literature for possible additional constraints on its existence. Three main type of observations can be used in this context: radial velocities, astrometric measurements, and high dynamic range single-pupil images. Before discussing the existing data in the literature, it is useful to convert the companion flux ratio into an estimated mass, using the evolutionary models of \citet{Baraffe98} for low-mass stars. Assuming an age of 200\,Myr for Fomalhaut \citep{DiFolco04}, and taking into account its distance of 7.7\,pc, a $K$-band magnitude of 6.1 for a close-in low-mass companion translates into a mass of $0.35\,M_{\odot}$. Wider companions would have larger masses to account for the reduced off-axis transmission of VLTI/VINCI, e.g., $0.48\,M_{\odot}$ at $0\farcs8$ (i.e., 6\,AU) where the interferometric transmission has dropped by a factor 2. This typically corresponds to spectral types in the range M1V--M3V.

Radial velocity monitoring of Fomalhaut was recently performed with HARPS by \citet{Lagrange09}, in the context of a radial velocity survey of early-type main sequence stars. Their observations of Fomalhaut span a 325 day period. Although Fomalhaut was found to be variable in radial velocity, no companion was detected with the following 68.2\% probability upper limits on companion masses as a function of orbital period: $3.6\,M_{\rm Jup}$ at 3 days, $3.2\,M_{\rm Jup}$ at 10 days, and $10\,M_{\rm Jup}$ at 100 days, the latter corresponding to a semi-major axis of 0.5\,AU. Since such planetary companions would produce flux ratios largely below 0.88\% in the $K$ band, this basically rules out any companion closer than 0.5\,AU as the source of the detected near-infrared excess.

An anomalous proper acceleration of 6.6\,mas/yr$^2$ has been observed in {\sc Hipparcos} astrometry towards Fomalhaut \citep{Chiang09}. This acceleration is marginally significant ($\sim2\sigma$), the Fomalhaut astrometry being otherwise stable. Such quasi-steady acceleration could be caused by a companion whose orbital period is longer than $\sim$\,3\,yr (i.e., semi-major axis $a\ge 2.6$\,AU). The required mass of this companion to explain the observed acceleration would be about $9\,M_{\rm Jup}$ at 2.6\,AU, and would increase to $50\,M_{\rm Jup}$ at 7\,AU, or $300\,M_{\rm Jup}$ (i.e., $0.29\,M_{\odot}$) at 15\,AU. The latter would produce a $K$-band flux ratio of about 0.6\% with respect to Fomalhaut. Companions of larger masses at the same orbital distances would be incompatible with the {\sc Hipparcos} data, while companions with orbital periods smaller than 3\,yr should have masses smaller than $9\,M_{\rm Jup}$ to be consistent with {\sc Hipparcos} observations. Because a companion with a 0.6\% flux ratio at an orbital semi-major axis of 15\,AU is mostly inconsistent with the VINCI data ($\chi^2_r \sim 1.8$, see Fig.~\ref{fig:chi2binary}), these astrometric measurements rule out the low-mass binary scenario as an explanation for the VINCI near-infrared excess up to orbital semi-major axes of about 15\,AU. At larger distances, the mass of a bound companion compatible with the {\sc Hipparcos} acceleration would become marginally consistent with the VINCI near infrared excess.

High contrast imaging with large diffraction-limited telescopes can be used put complementary constraints on low-mass companions at larger angular distances. In particular, the HST/ACS observations of Fomalhaut by \citet{Kalas05,Kalas08} show no low-mass companions at the sensitivity limit of the instrument, except for the planetary-mass companion at 119\,AU (well outside the VINCI field). According to \citet{Kalas02}, the ACS instrument can detect magnitude differences up to 11 at $0\farcs9$ (size of the Lyot occulting spot) for the F606W filter. Assuming a similar dynamic range in the F814W band, this gives a lower limit $I=12.0$ for the absolute magnitude of the companion in the $I$ band, which roughly corresponds to a $0.08\,M_{\odot}$ object at 200\,Myr and 7.7\,pc following the \citet{Baraffe98} models. This object would be much too faint to explain the $K$-band excess derived from VINCI observations, so that the effect of a companion located at an angular distance larger than $0\farcs9$ can be ruled out. For a low eccentricity orbit seen face-on, this corresponds to an orbital semi-major axis of 7\,AU. However, assuming the orbital plane of the potential companion to be coplanar with the circumstellar disk ($65\fdg9$ inclination) and taking into account the projection effect, the planet could be on a circular orbit with a semi-major axis as large as 14.3\,AU while still being hidden under the ACS coronagraphic mask for a 1.73\,yr time span, which corresponds to the interval between the two ACS images of \citet{Kalas05,Kalas08}. The HST/ACS images can thus rule out the binary scenario as the explanation of the VINCI near-infrared excess for semi-major axes of 14.3\,AU and above.

More recently, $M$ band observations with the AO-assisted Clio imager on the 6.5-m MMT have further constrained the range of allowed companion masses in the 5--40\,AU region around Fomalhaut \citep{Kenworthy09}. In particular, objects with masses larger than $13\,M_{\rm Jup}$ can be ruled out at distances between 8 and 40\,AU. Taking into account all these constraints, the presence of a bound low-mass companion around Fomalhaut, whatever its angular separation, cannot be at the origin of the near-infrared excess detected with VINCI. And since the probability to find a $K\sim6$ background object in the very small VINCI field is negligible, we are left with an extended source as the only plausible origin for the detected near-infrared excess.

    \subsection{Possible sources of extended emission}

Among the possible sources of extended emission in the close vicinity of Fomalhaut (mostly within $0\farcs8$), hot dust grains producing thermal emission in the near-infrared is the most straightforward scenario, since Fomalhaut is already known to be surrounded by large quantities of cold dust as well as by a warm dust population within 20\,AU \citep{Stapelfeldt04}. However, the thermal emission from hot dust grains is not the only possible source of extended near-infrared emission around Fomalhaut. In particular, non-thermal emission could be produced by free-free radiation from an ionized wind or a gaseous disk.

The possible presence of intense stellar winds around A-type main sequence stars has already been thoroughly discussed in \citet{Absil08}, showing that such stars have generally weak winds that are not expected to produce significant near-infrared excesses. The absence of detectable X-ray emission towards Fomalhaut \citep{Schmitt97} further suggests that a strong stellar wind cannot be at the origin of the observed $K$-band excess. Gaseous disks, on the other hand, are generally occurring in classical Be or Ae stars, a category of hot rapidly rotating stars close to the main sequence that exhibit prominent hydrogen emission lines. Although Fomalhaut does not rotate as fast as classical Ae/Be stars, and has a spectral type later than most emission-line stars, the possible presence of a gaseous disk cannot be ruled out \textit{a priori}. Since infrared excesses in such stars are generally correlated to the equivalent width of their H$\alpha$ emission line \citep[see e.g.][and references therein]{vanKerkwijk95}, we have searched the literature for possible evidence of H$\alpha$ emission around Fomalhaut. The H$\alpha$ photospheric absorption line profile of Fomalhaut was inspected in detail by \citet{Gardiner99} and \citet{Smalley02}. In particular, the observed H$\alpha$ line profile was fitted with a model Balmer profile to derive the effective stellar temperature, which is found by \citet{Smalley02} to be in agreement with the fundamental effective temperature derived by other methods. No peculiar feature is reported in the H$\alpha$ absorption line, including no obvious emission line inside it. Furthermore, the photometric index of the H$\alpha$ line was monitored between 1968 and 1974 by \citet{Dachs75}, showing no variability and no evidence of H$\alpha$ emission for Fomalhaut in 19 observations.

Finally, let us note that Fomalhaut is located well inside the Local Interstellar Bubble (LIB), so that the heating of interstellar material proposed to explain the presence of infrared excesses around shell stars or $\lambda$~Bo\"otis stars located about 100\,pc away \citep{Abt04,MartinezGalarza09} is unlikely to be at the origin of the observed near-infrared excess. Consequently, we consider the presence of hot dust in the inner planetary system of Fomalhaut as the only plausible explanation to the observed near-infrared excess, and further discuss this scenario in the next sections.


\section{Constraining the inner disk morphology} \label{sec:diskmodel}

We have previously derived in Section~\ref{sub:diskfit} the integrated flux ratio between the circumstellar disk and the stellar photosphere using a very simple brightness distribution for the disk: a uniform emission across the instrumental FOV. Accepting that an inner dust disk is indeed at the origin of the detected near-infrared excess, we discuss here the limitations to this preliminary analysis and investigate whether the morphology of the disk could be constrained based solely on our interferometric data set.

\begin{figure}
\plotone{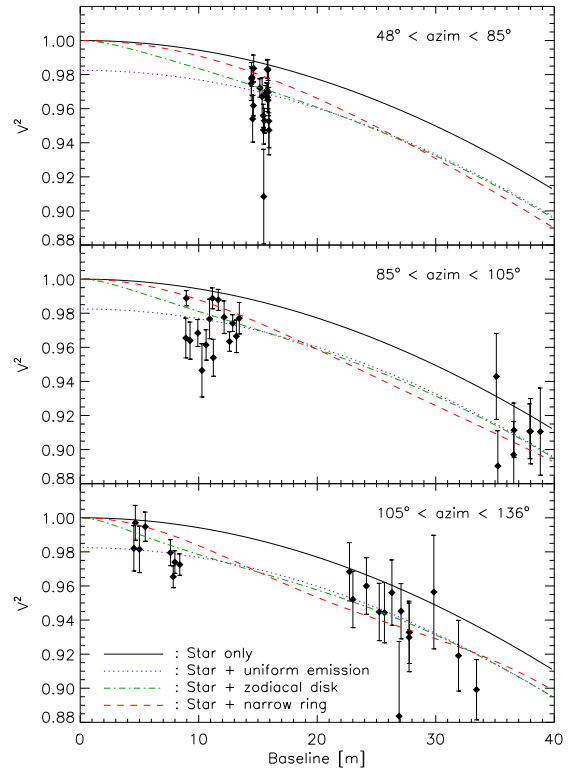}
\caption{Comparison of our data set with various disk models at short baselines. The data have been separated into three ranges of baseline azimuths to allow the comparison, since the model visibilities depend on both the baseline length and azimuth. In our models, the narrow ring is located at the sublimation radius of black body dust grains, assuming a sublimation temperature of 1700\,K, while the zodiacal disk follows the \citet{Kelsall98} model.}
\label{fig:models}
\end{figure}

As a first step, we have tried to use three very different morphologies for the circumstellar disk to reproduce the data set: a uniform emission extending across the whole field-of-view as in Section~\ref{sub:diskfit}, a narrow ring of dust\footnote{The ring width is forced to be much smaller than the interferometer's angular resolution so that it does not impact the visibility estimations.} located at the sublimation radius ($r_{\rm sub} \simeq 0.1$\,AU for a sublimation temperature $T_{\rm sub}=1700$\,K), and the zodiacal disk model of \citet{Kelsall98}, which is implemented in the \textsc{Zodipic} package\footnote{\textsc{Zodipic} is an IDL program developed by M.~Kuchner et al.\ for synthesizing images of exozodiacal clouds. It can be downloaded at {\tt http://asd.gsfc.nasa.gov/Marc.Kuchner/home.html}.}. Fig.~\ref{fig:models} illustrates the influence of the disk morphology on the squared visibility at short baselines. The small angular extent of the ring structure leads to significant oscillation in the squared visibility, whose frequency depends on the azimuth of the baseline since the disk is inclined. Smaller oscillations are visible at a similar frequency for the zodiacal disk model, because the inner rim dominates the $K$-band emission of the disk. Fitting these three models to our complete data set leads to similar results in terms of best-fit flux ratio and reduced $\chi^2$ (see Table~\ref{tab:chi2}), except for the ring structure which gives a larger best-fit contrast of $\sim1.3$\% with a slightly increased $\chi_r^2$. The larger best-fit contrast is mainly due to the fact that the ring is not fully resolved at the shortest baselines ($\lesssim 10$\,m), so that a larger disk emission is required to produce the same squared visibility deficit at those baselines. Note however that the best-fit contrast is only about $3\sigma$ above our original estimation, which indicates that, even with this extreme morphology, the final disk/star flux ratio does not heavily depend on the distribution of the circumstellar emission.

\begin{deluxetable}{cccc}
\tablecaption{Best-fit flux ratio and goodness of fit for three different disc models \label{tab:chi2}}
\tablewidth{0pt}
\tablehead{\colhead{} & \colhead{Uniform emission} & \colhead{Zodiacal disc} & \colhead{Narrow ring\tablenotemark{a}}}
\startdata
Flux ratio & 0.88\% & 0.85\% & 1.26\% \\
$\chi^2_r$ &  0.95  &  0.93  & 1.24
\enddata
\tablenotetext{a}{The ring is located at the dust sublimation radius ($\sim 0.1$\,AU).}
\end{deluxetable}

\begin{figure*}
\epsscale{1}
\plottwo{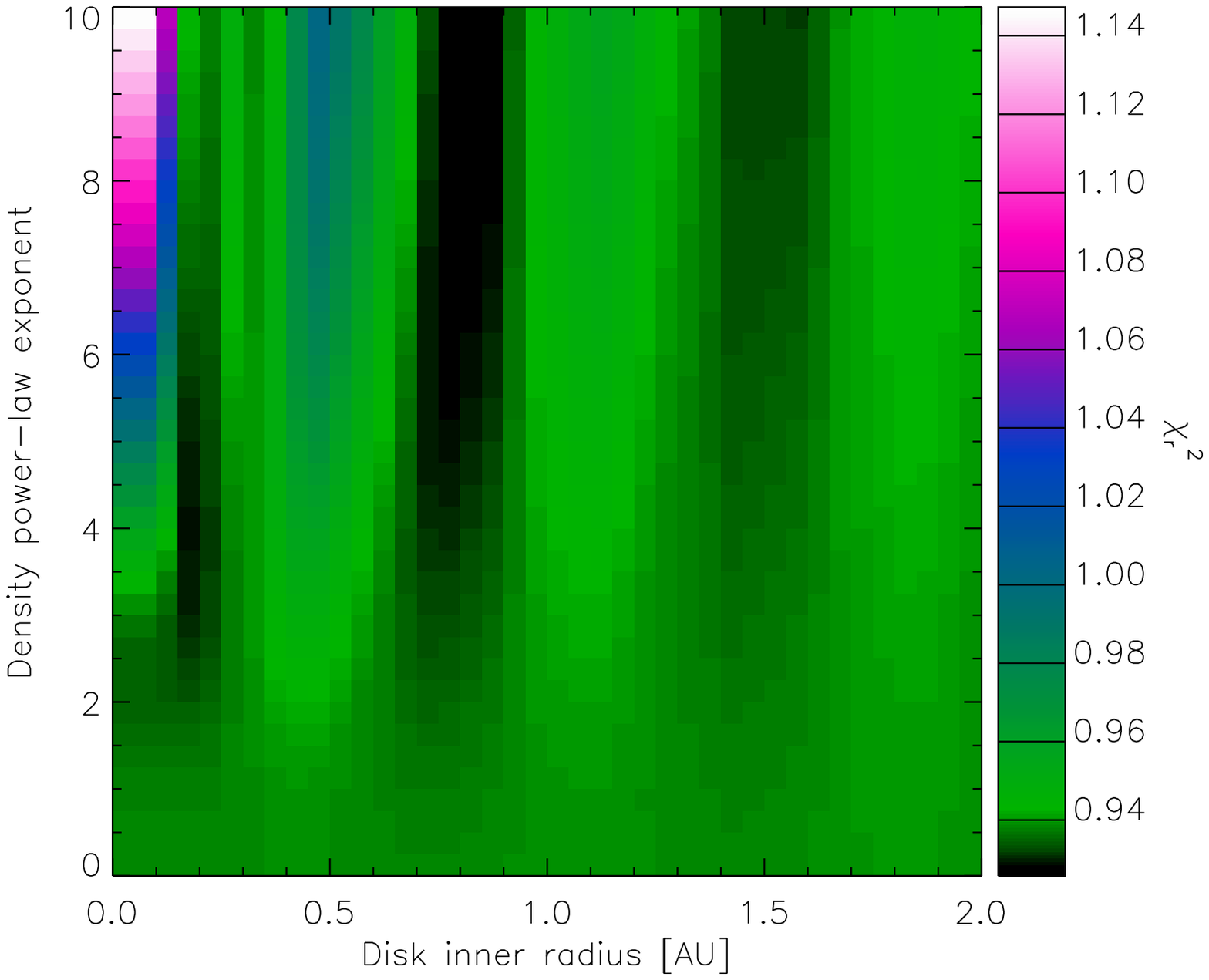}{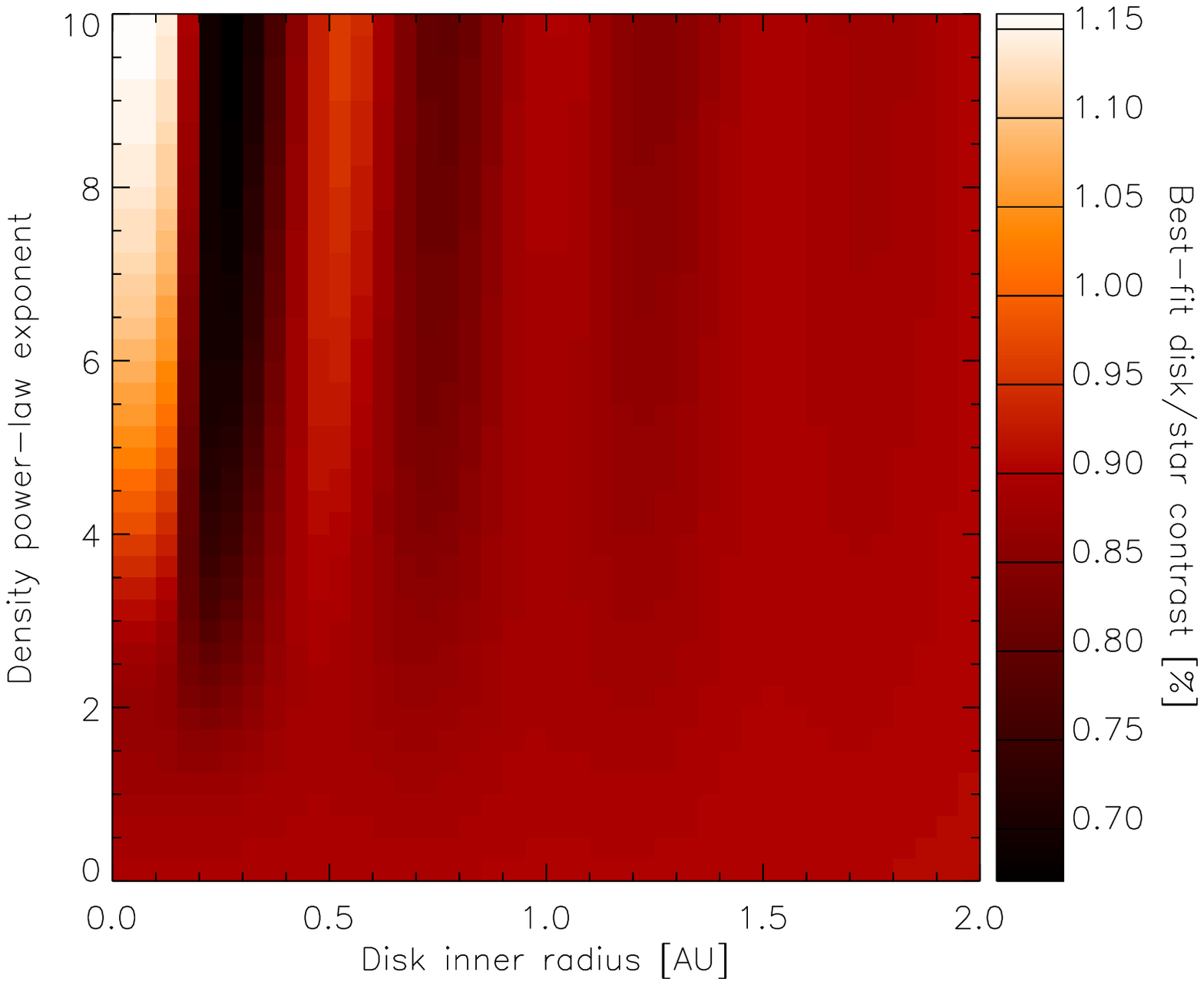}
\caption{\textit{Left}: Reduced chi square map for the fit of our simple geometrical disk toy model (see text for description), as a function of two free parameters: inner disk radius and density power-law exponent. The flux ratio between the disk and the star has been optimized to minimize the $\chi^2_r$ values. \textit{Right}: Optimum disk/star flux ratio associated to each disk model displayed on the $\chi^2$ map.}
\label{fig:chi2disk}
\end{figure*}

As a second step, we have generated images of a geometrically and optically thin debris disk toy model based on pure black body assumption (grain temperature proportional to $r^{-0.5}$, with $T=1700$\,K at 0.1\,AU), using a range of values for the two most important geometric parameters: the inner disk radius $r_{\rm in}$ and the exponent $\alpha$ of the power-law describing the density decrease as a function of distance ($n(r) \propto r^{-\alpha}$). The synthetic images only include thermal emission, which is expected to be largely dominant for the dust temperatures explored here. For each couple of parameters ($r_{\rm in}$, $\alpha$), we compute the visibility of the star-disk system at the relevant observing dates and adjust the disk/star flux ratio in order to minimize the $\chi^2$ distance between the modeled and observed visibilities. By exploring a range of possible values for the two parameters, we produce a $\chi^2$ map, which is displayed in Fig.~\ref{fig:chi2disk} together with the associated best-fit flux ratios. In these toy models, we have assumed a dust sublimation temperature of 1700\,K. All models with input inner radii smaller than the sublimation radius have their inner radius forced to $r_{\rm in}=0.1$\,AU and therefore give the same result.

The small range of reduced $\chi^2$ observed in the map of Fig.~\ref{fig:chi2disk} suggests that all models fit almost equally well our data set, so that the morphology of the circumstellar disk can not be meaningfully constrained. It must still be noted that models with very steep density profiles ($\alpha>6$) starting at the sublimation radius have a slightly larger $\chi_r^2$. These particular models roughly correspond to narrow rings of dust located at the sublimation radius, which do not reproduce very well the data at the shortest baselines (see Fig.~\ref{fig:models}). For dust rings with larger inner radii, the oscillation frequency in the visibility curve is increased (the disk is fully resolved at a shorter baseline) and several local minima are then obtained in the $\chi^2$ map, e.g.\ at $r_{\rm in}=0.2$\,AU, 0.8\,AU and 1.5\,AU. The solution at $r_{\rm in}=0.2$\,AU would correspond to a sublimation temperature of 1300\,K for black body dust grains. Conversely, models with flat density profiles ($\alpha<2$) give equal fit qualities whatever the inner radius, because their spatially extended brightness distribution produces almost no oscillation in the visibility curve (see green dash-dotted curve in Fig.~\ref{fig:models}). It must also be noted that, whatever the chosen morphology of the dust disk, the best-fit disk/star flux ratio remains similar to the value derived with the uniform emission model. The largest discrepancy is found for the models with steep density power-laws starting at the sublimation radius, which give a best-fit contrast of 1.16\% (i.e., about $2\sigma$ above our original estimation). This observation reinforces our statement that the disk morphology has a weak influence on the best-fit disk/star contrast.


\section{Discussion} \label{sec:discussion}

The discovery of hot dust grains in the inner planetary system of Fomalhaut cannot be regarded as a totally unexpected result. First, the data collected by \citet{Stapelfeldt04} with the MIPS and IRS instruments onboard Spitzer have already shown the presence of a warm excess emission at wavelengths between 17 and 34\,$\mu$m. MIPS imaging suggests that most of the 24\,$\mu$m excess is unresolved and originates from a $10\arcsec$ region around Fomalhaut, i.e., well within the main dust ring located at 140\,AU. Second, other A-type stars have already been shown to harbor similar near-infrared excesses that have been interpreted as the signature of hot dust \citep{Absil06,Absil08,Akeson09}. In all previous cases, the absence of significant excess emission at wavelengths around 10\,$\mu$m has suggested that the dust surface density distribution is very compact and does not follow a classical $r^{-1.5}$ power-law. Modeling these interferometric observations has still shown that realistic debris disk models can be found to fit the observed near-infrared excesses of a few percent around A-type stars without producing too large amounts of excess at longer wavelengths, that would have been noticed in mid-infrared spectro-photometric measurements.

It is not our purpose here to thoroughly study the properties of the Fomalhaut inner disk, as this will be done in a companion paper, where we include additional interferometric data collected at 10\,$\mu$m with the Keck Interferometer Nuller. It is nevertheless useful to check whether standard disk models could both reproduce the measured $K$-band excess and be consistent with photometric data at longer wavelengths. For that, we have used our debris disk toy model based on pure black body grains, assuming an $r^{-1.5}$ surface density power law from the dust sublimation radius up to 140\,AU. In order to reproduce the measured $K$-band excess, this disk must have a bolometric luminosity ratio $L_{\rm disk}/L_{\ast} \sim 5\times 10^{-4}$ with respect to the stellar photosphere. The excesses produced at 17.5 and 24\,$\mu$m would then be of 0.38 and 0.22\,Jy respectively, which would correspond to the measured IRS ``on-star'' excess of $0.4 \pm 0.2$\,Jy at 17.5\,$\mu$m, while being fully within the measured MIPS ``on-star'' excess of $0.6 \pm 0.2$\,Jy at 24\,$\mu$m \citep{Stapelfeldt04}. The predicted flux in the ``extended'' disk located outside $10\arcsec$, amounting to 0.06\,Jy, would also be within the MIPS PSF-subtracted estimation of $0.13 \pm 0.02$\,Jy for the outer disk regions. Our model is also consistent with published photometric measurements in the mid-infrared (including an IRAS 12\,$\mu$m observation), which are generally of poor accuracy. We will however show in a companion paper that this simple model cannot reproduce interferometric data obtained in the mid-infrared with the Keck interferometer, as well as unpublished IRS data in the 10--18\,$\mu$m region. This suggests that the disk cannot be continuous under the $r^{-1.5}$ power law assumption, and therefore cannot be directly connected with the outer belt using classical models (e.g., based on Poynting-Robertson drag).


\section{Conclusion}

In this paper, we have re-analyzed published and unpublished archival data from the VLTI/VINCI instrument in search for near-infrared circumstellar emission around the young A4V star Fomalhaut. The large number of visibility measurements at short baselines (146 individual measurements with projected baselines smaller than 60\,m), and the good accuracy of individual data points (median relative error bar of 2\%), have allowed us to directly detect the presence of resolved circumstellar emission at a level of $0.88\% \pm 0.12\%$ with respect to the stellar photosphere in the $K$ band. If this excess emission was produced by a dust disk similar to the solar zodiacal disk, its surface density would be about 5000 times as large as in the solar system. Our analysis of the whole VINCI data set also provides an updated mean limb-darkened diameter $\theta_{\rm LD}=2.223 \pm 0.022$\,mas for the stellar photosphere, which is predicted to have an apparent oblateness of 1.021. This measurement does not significantly differ from the previous estimation by \citet{DiFolco04}, which did not take into account the presence of circumstellar emission.

We attempted to constrain the morphology of the circumstellar emission source with our interferometric data set, but failed to discriminate between a point-like source and an extended source, mostly due to the inappropriate observing strategy of this heterogeneous data set. Additional constraints obtained with complementary observing techniques have thus been used to determine the nature of the excess emission, showing that the presence of a point-like source within the interferometric field-of-view (either a bound companion or a background object) is very unlikely to reproduce the observed $K$-band excess. Non-thermal extended sources are also shown to be mostly inconsistent with other observations. We therefore consider that a hot dust population located mostly within the first 6\,AU around Fomalhaut is the most likely explanation for the observed $K$-band excess.

In a companion paper, we will introduce further constraints on the inner dust disk thanks to high dynamic range mid-infrared observations with the Keck Interferometer Nuller, and complement these new constraints with various spectro-photometric measurements in order to investigate the nature and distribution of the dust grains.


\acknowledgments

O.A.\ acknowledges the financial support from the European Commission's Sixth Framework Program as a Marie Curie Intra-European Fellow (EIF) while at LAOG, and from a F.R.S.-FNRS Postdoctoral Fellowship while at IAGL. This research was partly funded by the International Space Science Institute (``Exozodiacal Dust Disks and Darwin'' working group) and by an EGIDE/PHC Procope programme (\#17843ZE).



{\it Facilities:} \facility{VLTI(VINCI)}



\appendix

\section{Oblate limb-darkened photospheric model} \label{app:photosphere}

To properly fit our interferometric data, which are taken at various baseline azimuths, we must take into account the influence of stellar oblateness due to the rapid rotation of Fomalhaut \citep[$v \sin i = 93$\,km\,s$^{-1}$,][]{Royer07}. Following the discussion of \citet{Absil08}, we compute the apparent oblateness $\rho$ of the Fomalhaut photosphere, defined as the ratio of the major and minor apparent radii ($\rho = R_a/R_b$), using the following equation:
\begin{equation}
\rho \simeq \left( \frac{(v \sin i)^2 R_m}{4GM_{\ast}} + \sqrt{1 + \left( \frac{(v \sin i)^2 R_m}{4GM_{\ast}} \right)^2} \right)^2 \, , \label{eq:oblateness}
\end{equation}
where $R_m=(R_aR_b)^{1/2}$ is the geometric mean between the major and minor apparent radii. Taking for $R_m$ the stellar radius derived by \citet{DiFolco04} based on long-baseline interferometric measurements ($R_m = 1.840 \, R_{\odot}$), and with $M_{\ast}=2.0 \, M_{\odot}$ \citep{DiFolco04}, the apparent oblateness of the photosphere is $\rho = 1.021$. The orientation of the photosphere on the plane of the sky has recently been constrained by spectrally dispersed interferometric measurements using VLTI/AMBER \citep{Lebouquin09}, showing a perfect match between the position angle of its major axis (${\rm PA_{star}}=156\degr \pm 3 \degr$) and that of the debris ring \citep[${\rm PA_{disk}}=156\fdg0\pm 0\fdg3$,][]{Kalas05}. We will further assume that the photospheric limb darkening can be approximated by a linear law and use the $K$-band limb-darkening coefficient tabulated by \citet{Claret00} for an A4\,V star ($u_K=0.19$). With these assumptions, we are left with only one parameter to fit the interferometric data set: the mean limb-darkened angular diameter $\theta_{\rm LD}$.

\section{Importance of short baselines for debris disk detection} \label{app:shortbase}

Another representation of the fit result described in Section~\ref{sub:diskfit} and Fig.~\ref{fig:fitdisk1} is proposed in Fig.~\ref{fig:fitdisk2}, where we have separately displayed the data collected with the two shortest baseline configurations (E0--G0 and D0--H0). The squared visibility associated with the best-fit stellar photosphere model (fitted on the whole data set) is depicted by a blue curve, whose thickness represents the uncertainty on the squared visibility resulting from both the statistical and systematic error bars on its diameter. The presence of the circumstellar disk is revealed by the fact that all data points are located significantly below the expected stellar visibility. Even though the circumstellar disk is most evident at the shortest baseline (E0--G0), where the stellar photosphere is completely unresolved, the intermediate D0--H0 baseline alone would be sufficient to conclude on its presence, although with a larger error bar on the final disk/star contrast. This figure clearly illustrates the importance of short baselines in the detection of circumstellar disks around main sequence stars, as the visibilities at longer baselines are much more affected by the uncertainties on the stellar photospheric model. In particular, we stress that, thanks to the very short baselines used here, modifications in the fixed stellar limb-darkening and oblateness parameters would not significantly change the final estimation of the disk/star contrast.




\bibliographystyle{apj} 
\bibliography{Fomalhaut} 

\end{document}